\documentclass[12pt]{iopart}
\usepackage{graphicx}
\usepackage{amssymb}
\usepackage{epsfig}
\usepackage{subfig}

\newcommand{\ud}{\mathrm{d}}

\begin{document}
\title{Inflation of the screening length induced by Bjerrum pairs}
\author{Jos Zwanikken and Ren\'{e} van Roij}
\address{Institute for Theoretical Physics, Utrecht
University, Leuvenlaan 4, 3584 CE Utrecht, The Netherlands}

\begin{abstract}
Within a modified Poisson-Boltzmann theory we study the effect of
Bjerrum pairs on the typical length scale $1/\bar{\kappa}$ over
which electric fields are screened in electrolyte solutions,
taking into account a simple association-dissociation equilibrium
between free ions and Bjerrum pairs. At low densities of Bjerrum
pairs, this length scale is well approximated by the Debye length
$1/\kappa\propto 1/\sqrt{\rho_\mathrm{s}}$, with $\rho_\mathrm{s}$
the free ion density. At high densities of Bjerrum pairs, however,
we find $1/\bar{\kappa}\propto \sqrt{\rho_\mathrm{s}}$ which is
significantly larger than $1/\kappa$ due to the enhanced effective
permittivity of the electrolyte, caused by the polarization of
Bjerrum pairs. We argue that this mechanism may explain the
recently observed anomalously large colloid-free zones between an
oil-dispersed colloidal crystal and a colloidal monolayer at the
oil-water interface.
\end{abstract}


\maketitle

\section{Introduction}\label{sec:Bjerrum pairs}

Phase separation and criticality in electrolyte solutions,
following from Debye-H\"uckel (DH) theory \cite{Debye}, was
extensively studied by Fisher and Levin \cite{Fisher}. The
calculated critical density and temperature differed by only
$\sim$10 \% from the results of Monte Carlo simulations by
Panagiotopoulos \cite{Panagiotopoulos}, showing DH-theory to be a
reliable basis to describe some basic features of electrolyte
solutions. Fisher and Levin extended the original DH-theory by
inclusion of \textsl{Bjerrum pairs}, following the ideas of
Bjerrum \cite{Bjerrum} that plus and minus ions can form neutral
pairs. These Bjerrum pairs or \textsl{dipoles} are considered as a
separate particle component, and reduce, within the restricted
primitive model, the total number of free ions. The results of the
extended DH-theory agreed remarkably with simulation results,
especially when couplings between the dipoles and the ions, as
well as hard-core repulsions were taken into account
\cite{Fisher}. Here, we apply the same ideas of Bjerrum to
describe screening effects in low-dielectric solvents ('oils') by
means of a modified Poisson-Boltzmann theory. In these solutions
we expect strong correlations between the ions, since the energy
gain of bringing two oppositely charged ions at contact can exceed
the thermal energy considerably.

We will consider systems of free ions and dipolar particles,
similar to the study of e.g. \cite{Abraskin}, and find that
electric fields are screened over a typical length scale
$1/\bar{\kappa}$ that can be significantly larger than the Debye
length $1/\kappa$ (based on the ionic strength), at least for
large dipole densities. We predict that these densities are to be
expected in low-dielectric solvents by considering a simple
association-dissociation equilibrium between the free and bound
ions \cite{Camp}. Recently, strong electrostatic repulsions were
observed by Leunissen et al. \cite{Leunissen} in low-dielectric
solvents ($4\lesssim\epsilon\lesssim10$), sometimes extending over
a length scale beyond $100$ $\mu$m. The Debye length
$1/\kappa^{-1}$ was found to be only $\sim 4$ $\mu$m, calculated
from conductivity measurements (the density of free ions). The
analysis presented in this paper provides a possible explanation
and a reason for the quantitative differences between these
experiments \cite{Leunissen} and the theory in \cite{Zwanikken},
where Bjerrum pairs were not taken into account.

In section \ref{subsec:Equilibrium constants} a reaction
equilibrium between free and paired ions will be discussed.
Parameter space will be divided into regions where bound ions
outnumber the free ions, and vice versa, as was already
calculated, e.g. in \cite{Camp}. In section
\ref{sec:EffectiveScreening} the effect of the dipole density on
the screening length will be analyzed. First we extend
Gouy-Chapman theory \cite{Gouy} by including dipoles and calculate
the {\sl effective} screening length $1/\bar{\kappa}$, similar to
the calculations in \cite{Abraskin}. Remarkably, we find
$1/\bar{\kappa}\propto \sqrt{\rho_\mathrm{s}}$ at high dipole
densities, in contrast to $1/\kappa\propto
1/\sqrt{\rho_\mathrm{s}}$, where $\rho_\mathrm{s}$ is the density
of free ions. Finally we review and extend the theory presented in
\cite{Zwanikken} and predict a larger effective screening length
due to dipoles. The considered densities of free ions are well
below the critical value, i.e. phase equilibrium is not
considered, even though the temperatures of interest are close to
the critical temperature.

\section{Bjerrum pairs}\label{subsec:Equilibrium
constants}

\begin{figure}[!ht]
\centering
\includegraphics[width = 12cm]{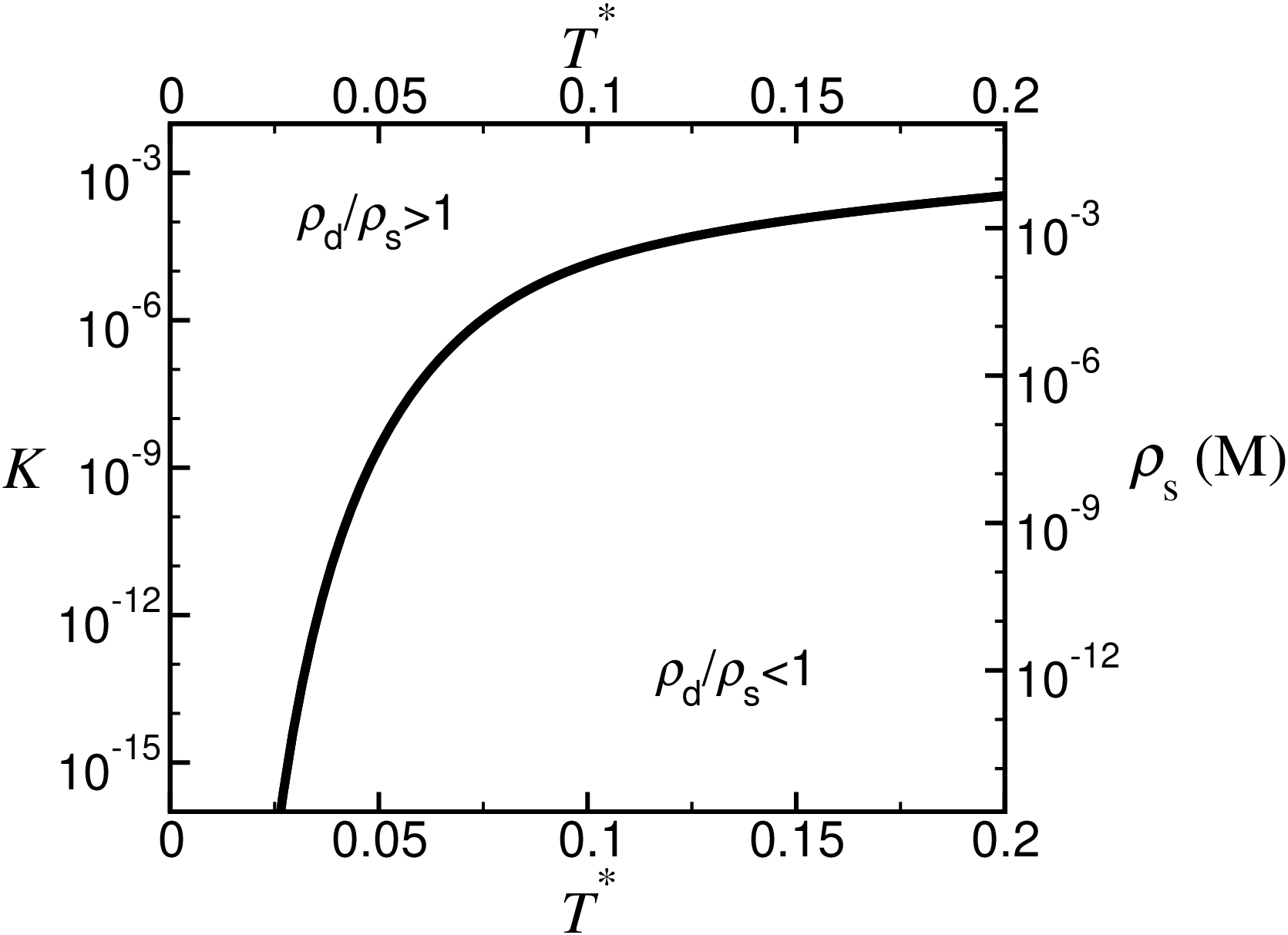}
\caption{\small The dimensionless equilibrium constant
$K\sigma^3=K$ as a function of the dimensionless temperature
$T^*=\sigma/\lambda_B$. The line also marks the points where the
density of paired ions equals the density of free cations/anions
for given concentration of free ions (see right vertical axis),
for an ionic diameter $\sigma=0.5$ nm.} \label{fig:Kconstant4}
\end{figure}

First we consider a three-dimensional bulk electrolyte of
monovalent cations and anions, at a total density of
$2\rho_\mathrm{tot}$. The ions may form pairs or remain free; the
number density of dipolar Bjerrum pairs is $\rho_\mathrm{d}$ and
the number densities of free ions are
$\rho_+=\rho_-=\rho_\mathrm{s}$. In terms of dimensionless
densities $\eta_\mathrm{x} = \rho_\mathrm{x} \sigma^3$, where
$\sigma$ is the common diameter of the ions, the total density
$\eta_\mathrm{tot}$ of ions (of one type) is
\begin{equation}\label{fml:ParticleConserve}
\eta_\mathrm{tot} = \eta_\mathrm{d} + \eta_\mathrm{s}.
\end{equation}
The strength of the electrostatic interactions in the solvent is
reflected by the Bjerrum length, which is the length at which the
bare Coulomb interaction between two monovalent ions is exactly 1
$k_\mathrm{B}T$,
\begin{equation}
\lambda_\mathrm{B} = \frac{ e^2 }{ 4\pi \epsilon k_\mathrm{B}T },
\end{equation}
where $e$ is the elementary charge, and $\epsilon$ the dielectric
constant of the medium. For water at room temperature, this length
is only 0.71 nm, for apolar solvents it measures up to several
tens of nanometers, and in vacuum it is $\sim57$ nm. The Coulomb
interaction between two ions can hence be written in terms of the
Bjerrum length,
\begin{equation}
\frac{V_\mathrm{C} (r)}{k_\mathrm{B}T} =
\pm\frac{\lambda_\mathrm{B}}{r} \equiv\pm\frac{1}{l},
\end{equation}
where the $+,-$ refer to equal charged particles, and oppositely
charged particles, respectively, $r$ is the distance between the
particles, and $l$ is a dimensionless distance. The dimensionless
equilibrium constant $K$ of the reaction of free ions that bind
into paired ions is defined by
\begin{equation}\label{fml:ReactionConstantBjerrum}
K = \frac{\eta_+\eta_-}{\eta_\mathrm{d}} =
\frac{\eta_\mathrm{s}^2}{\eta_\mathrm{d}}=
\frac{\sigma^3}{\Lambda^3} \exp\Big(\frac{\Delta G}{k_BT}\Big),
\end{equation}
where $\Lambda$ is the ionic Debroglie wavelength, and where
$\Delta G$ is the free energy of a bound pair of ions (being
separated by a distance $\sigma < r <\lambda_\mathrm{B}$), with an
associated Coulombic binding energy
$V_\mathrm{C}=-k_\mathrm{B}T/l$. It will be convenient to
introduce the dimensionless temperature
$T^*=\sigma/\lambda_\mathrm{B}$, such that $K^{-1}$ can be
expressed in terms of an internal partition function
\begin{equation} \label{fml:EqConst}
K^{-1} = 4\pi \Big(\frac{1}{T^*}\Big{)}^3 \int \limits_{T^*}^1 \ud
l\ l^2\ \exp\bigg(\frac1{l}\bigg),
\end{equation}
as already postulated by Bjerrum \cite{Bjerrum}. It can easily be
checked that $T^*\simeq 1$ for typical ions such as Na$^+$ and
Cl$^-$ in water at room temperature, and $T^*\lesssim 0.2$ in oils
with $\epsilon\lesssim15$. Figure \ref{fig:Kconstant4} shows the
relation between the equilibrium constant $K$ and the
dimensionless temperature $T^*$. With increasing temperature $K$
rises, thereby lowering the tendency to form pairs according to
(\ref{fml:ReactionConstantBjerrum}). Using
(\ref{fml:ParticleConserve}) this can be further quantified by
relating $\eta_\mathrm{s}$ to $\eta_\mathrm{tot}$ as
\begin{equation}
\eta_\mathrm{tot} = \eta_\mathrm{s} + \frac{\eta_\mathrm{s}^2}{K},
\end{equation}
which yields $\eta_\mathrm{s}$ and $\eta_\mathrm{d}$ as a function
of $\eta_\mathrm{tot}$ and $K(T^*)$ as
\begin{equation}\label{fml:RhoRatio}
\frac{\eta_\mathrm{s}}{\eta_\mathrm{tot}} =
\frac{K}{2\eta_\mathrm{tot}}\Big(\sqrt{1+\frac{4\eta_\mathrm{tot}}{K}}-1\Big)
= 1-\frac{\eta_\mathrm{d}}{\eta_\mathrm{tot}}.
\end{equation}


\begin{figure}[!ht]
\centering
\includegraphics[width = 7.1cm]{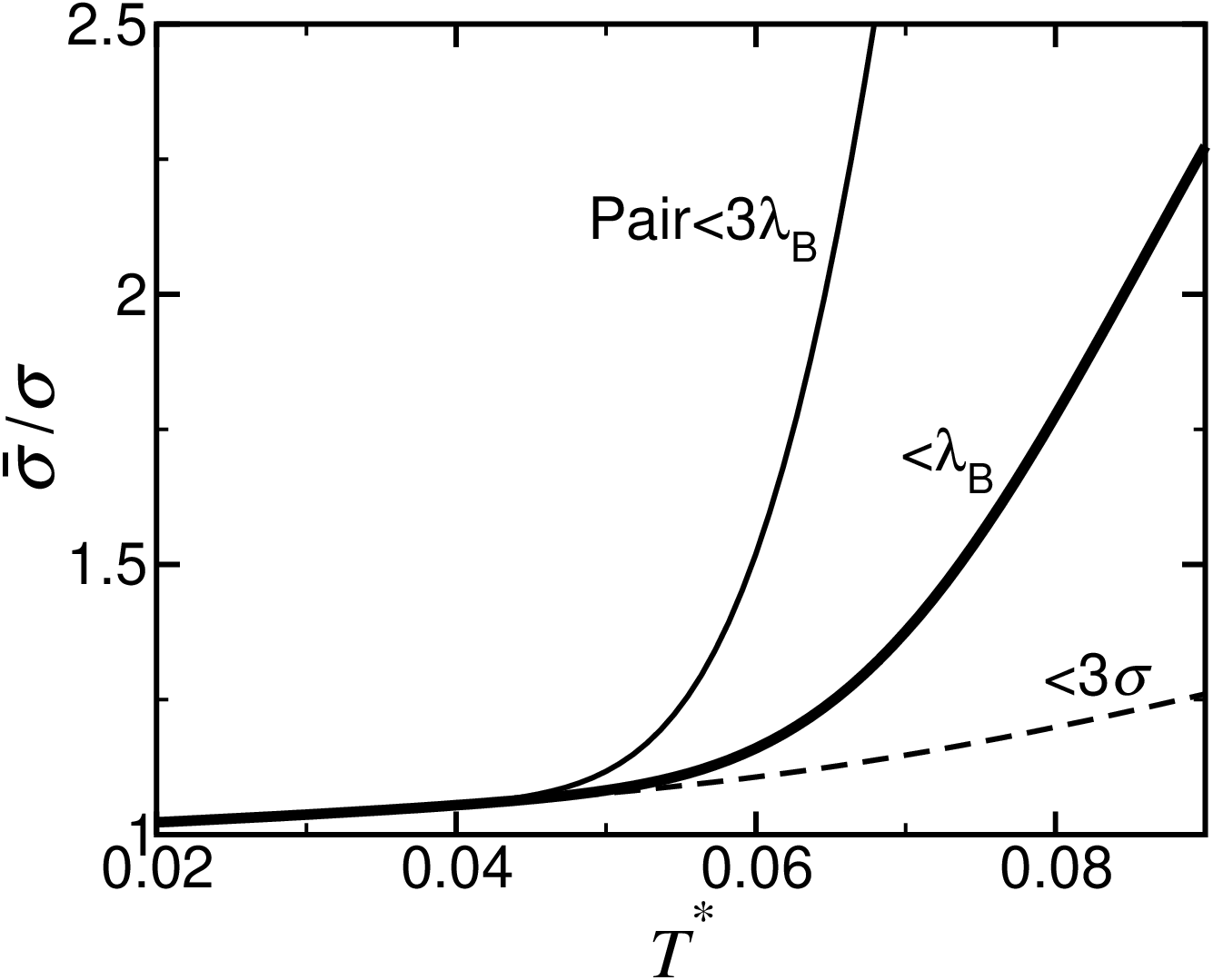}
\caption{\small The mean ion separation of the Bjerrum pairs
$\bar{\sigma}$ (in units of the ionic diameter $\sigma$) as a
function of the dimensionless temperature $T^*=\sigma/\lambda_B$,
for several definitions of the maximum ion separation that is
still called a Bjerrum pair. At low temperatures $T^*\lesssim
0.05$ the separation does not depend on the precise definition,
because the deep potential well of $>20$ $k_\mathrm{B}T$ at
contact dominates the partition sum of the pair.}
\label{fig:rMEAN}
\end{figure}

From (\ref{fml:ReactionConstantBjerrum}) it can be easily seen
that if $K = \eta_\mathrm{s}$, then
$\eta_\mathrm{s}=\eta_\mathrm{d}$. If the left vertical axis of
figure \ref{fig:Kconstant4} is read as the dimensionless density
$\eta_\mathrm{s}$ of free ions, the curve $K(T^*)$ thus separates
the parameter regime where dipoles dominate
($\eta_\mathrm{d}/\eta_\mathrm{s}>1$), from the regime where free
ions dominate ($\eta_\mathrm{d}/\eta_\mathrm{s}<1$). The right
vertical axis converts the corresponding $\eta_\mathrm{s} =
K(T^*)$ to the molar density $\rho_\mathrm{s}$ for the typical
choice $\sigma = 0.5$ nm. We have already seen that $T^*\simeq 1$
for typical ions such as Na$^+$ and Cl$^-$ in water at room
temperature, and that $T^*\lesssim0.2$ in oils with $\epsilon
\lesssim 15$. Figure \ref{fig:Kconstant4} therefore illustrates,
for instance, that an electrolyte with a millimolar ionic
strength, $\rho_\mathrm{s} \simeq 1$ mM, is dominated by free ions
for $T^*\gtrsim 0.15$ and by dipoles for $T^*\lesssim 0.15$. For
nanomolar concentrations, $\rho_\mathrm{s}=1$ nM, the crossover is
at $T^*\simeq 0.05$. For later reference we also consider the mean
separation $\bar{\sigma}$ between the ions in a pair. Assigning a
statistical weight $\propto\exp(\lambda_\mathrm{B}/r)$ to a pair
at separation $r$, one finds
\begin{equation} \label{fml:MeanSigma}
\bar{\sigma}^2 = \sigma^2\frac{\langle l^2 \rangle}{{T^*}^2} =
\sigma^2 K 4\pi \Big(\frac{1}{T^*}\Big{)}^5 \int \limits_{T^*}^1
\ud l\ l^4\ \exp\bigg(\frac1{l}\bigg),
\end{equation}
which can straightforwardly be evaluated numerically. The result
is shown in figure \ref{fig:rMEAN}, as a function of $T^*$, for
several upper bounds of the integration domain to check the
dependence of $\bar{\sigma}$ and $K$ on the definition of a
Bjerrum pair, being two oppositely charged particles separated by
a distance $\lesssim \lambda_\mathrm{B}$. At low temperatures
$T^*\lesssim 0.05$ the values do not depend on the precise
definition, because the deep potential well of $>20$
$k_\mathrm{B}T$ at contact dominates the probability distribution.

\section{Effective screening length}\label{sec:EffectiveScreening}

We now consider a system of monovalent ions near a charged plate
at $z=0$, where the $z$-axis is perpendicular to the plate, and
the ion density is $\rho_\mathrm{s}$ at $z\rightarrow\infty$. We
expect that the charge of the plate is screened by an oppositely
charged ionic cloud of net charge, generated by the ions. By the
Poisson-Boltzmann equation, solved by Gouy and Chapman
\cite{Gouy}, the typical width of this cloud (the double layer)
can be found. This length scale is also known as the Debye length
$1/\kappa = 1/\sqrt{8\pi\lambda_\mathrm{B}\rho_\mathrm{s}}$. We
now extend Gouy-Chapman theory \cite{Gouy} by the inclusion of an
additional particle species of dipoles with number density
$\rho_\mathrm{d}$ in the bulk far from the plate. Following the
derivation of \cite{Abraskin}, we find the Poisson-Boltzmann
equation for $z>0$
\begin{equation}\label{fml:PBAbrashkin}
\phi''(z) = \kappa^2 \sinh \phi(z) - \kappa^2
\frac{\rho_\mathrm{d}}{2\rho_\mathrm{s}} \bar{\sigma}
\frac{\ud}{\ud z}\big[\mathcal{G}(\bar{\sigma}\phi'(z))\big],
\end{equation}
where a prime denotes a derivative with respect to $z$, and
$/k_\mathrm{B}T\phi(z)/e$ is the electrostatic potential. The
function $\mathcal{G}(u) = \cosh(u)/u - \sinh(u)/u^2$ can
accurately be approximated by a first order expansion
$\mathcal{G}(u)= \frac13 u + \mathcal{O}(u^2)$, since the mean
separation $\bar{\sigma}=\mathcal{O}(1)$ nm, and the electric
field $\phi'(z)=\mathcal{O}(1)$ $\mu$m$^{-1}$ for the systems of
our interest. The PB-equation then reduces to
\begin{equation}
\phi''(z) = \bar{\kappa}^2 \sinh \phi(z),
\end{equation}
where
\begin{equation}\label{fml:ReducedKappa}
\bar{\kappa}^2 = \frac{\kappa^2}{\alpha\kappa^2
 + 1},
\end{equation}
with $\alpha \equiv
\rho_\mathrm{d}\bar{\sigma}^2/6\rho_\mathrm{s}$. The presence of
dipoles thus increases the screening length significantly as soon
as $\alpha\kappa^2 = \mathcal{O}(1)$, which can only be obtained
at high ionic strength in low dielectric media, such that
$\rho_\mathrm{d}/\rho_\mathrm{s}$ is large. Equivalently, one can
also consider the dielectric constant to be effectively changed by
the presence of the dipoles. Writing $\bar{\kappa} = 8\pi e^2
\rho_\mathrm{s}/(\bar{\epsilon}k_\mathrm{B}T)$, with
$\bar{\epsilon}$ the effective dielectric constant gives with
(\ref{fml:ReducedKappa}) that
\begin{equation}\label{fml:IncreasedEpsilon}
\bar{\epsilon} = \epsilon + \frac{4\pi\bar{\sigma}^2
\lambda_\mathrm{B}^\mathrm{vac}\rho_\mathrm{d}}{3},
\end{equation}
where $\lambda_\mathrm{B}^\mathrm{vac}$ is the Bjerrum length in
vacuum. The molar density of pairs has to be large enough for a
significant change in the effective dielectric constant. For
typical ion diameters of a few \AA ngstr\"om one needs
$\rho_\mathrm{d}\gtrsim 10$ mM for $\bar{\epsilon}\gtrsim
2\epsilon$.

\begin{figure}[!ht]
\centering

\caption{\small The effective screening length as a function of
the Debye length $\kappa^{-1}=(8 \pi \lambda_\mathrm{B}
\rho_\mathrm{s})^{-\frac12}$ (a), and dimensionless temperature
$T^*$ (b) on the basis of (\ref{fml:ReducedKappa2}).}

\subfloat[The (dimensionless) effective screening length as a
function of the Debye length $\kappa^{-1}$ showing the asymptotic
regimes $\bar{\kappa}^{-1}=\kappa^{-1}$ and $\bar{\kappa}^{-1}=
\kappa A^2$, separated by a minimum at $\kappa^{-1} = 2^\frac14 A$
(see text).]{\includegraphics[width = 5.5cm]{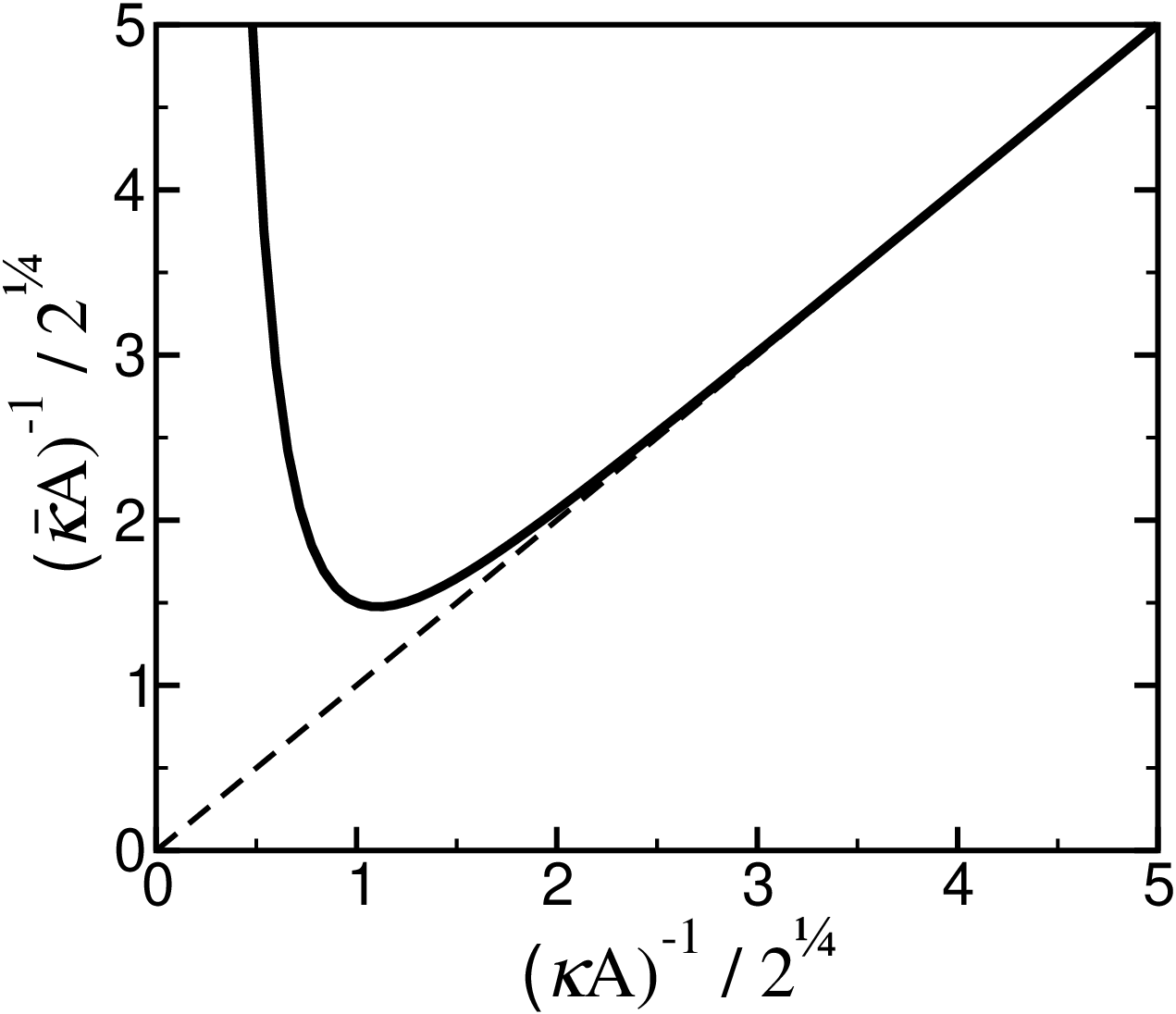}}
\hspace{5mm} \subfloat[The effective screening length (in units of
the Debye length) as a function of dimensionless temperature $T^*$
for several dipole densities $\rho_\mathrm{d}=0.1$, 1, 10 M, at a
mean separation and particle diameter $\bar{\sigma}=\sigma = 0.5$
nm. The upper axis shows the conversion from $T^*$ to $\epsilon$
at $\sigma = 0.5$ nm. The effective screening length can be up to
a factor $\mathcal{O}(10)$ higher than the Debye length
$\kappa^{-1}$, at high dipole densities.]{\includegraphics[width =
7cm]{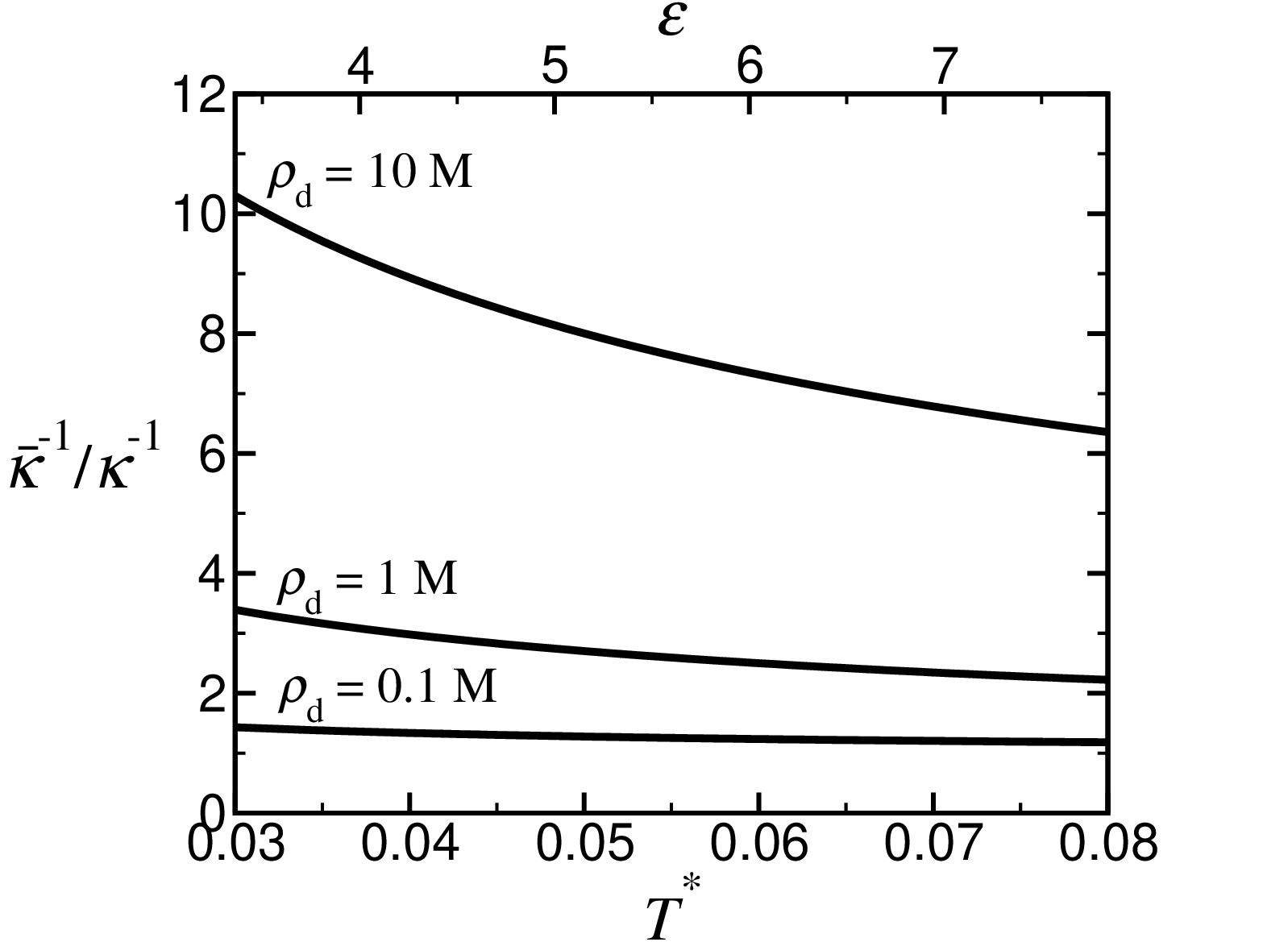}}

\label{fig:ReducedKappa}
\end{figure}

In (\ref{fml:PBAbrashkin})-(\ref{fml:IncreasedEpsilon}) we treated
$\rho_\mathrm{s}$ and $\rho_\mathrm{d}$ as independent densities,
whereas in (\ref{fml:ReactionConstantBjerrum}) we related them
directly through an equilibrium reaction. Using
(\ref{fml:ReactionConstantBjerrum}) we find $\alpha =
\eta_\mathrm{s}\bar{\sigma}^2/(6 K)$ such that the limit $\alpha
\kappa^2 \gg 1$ gives
\begin{equation}\label{fml:kappaDep}
\bar{\kappa}^{-1} = \sqrt{\alpha} \propto \sqrt{\rho_\mathrm{s}},
\end{equation}
which is a remarkable dependency, since in the absence of dipoles
the Debye length scales as $\kappa^{-1}\propto
1/\sqrt{\rho_\mathrm{s}}$. The full dependence of
$\bar{\kappa}^{-1}$ on $\kappa^{-1}$ follows from
(\ref{fml:ReducedKappa}) and (\ref{fml:ReactionConstantBjerrum})
as
\begin{equation}\label{fml:ReducedKappa2}
\bar{\kappa}^{-1} = \kappa^{-1}\sqrt{(A\kappa)^4 + 1} \approx
\left\{
\begin{array}{ll}
\displaystyle \frac{A^2}{\kappa^{-1}} &, \kappa^{-1} \ll A;\\
\kappa^{-1} &, \kappa^{-1} \gg A,
\end{array}
\right.
\end{equation}
with $A^4=T^*\sigma^2\bar{\sigma}^2/(48\pi K)$. Relation
(\ref{fml:ReducedKappa2}) is plotted, in a conveniently scaled
fashion, in figure \ref{fig:ReducedKappa}a, revealing a minimum at
$\kappa^{-1} = 2^\frac14 A$ that separates the dilute limit (where
$\kappa \simeq \bar{\kappa}$) from the dense limit (where
$\bar{\kappa}^{-1} \sim A^2\kappa$). In figure
\ref{fig:ReducedKappa}b the ratio $\bar{\kappa}^{-1}/\kappa^{-1}$,
which equals $\sqrt{(A\kappa)^4 +1}$ from
(\ref{fml:ReducedKappa2}), is plotted as a function of $T^*$ for
$\rho_\mathrm{d}=0.1$, 1, 10 M, setting $\bar{\sigma}=\sigma =
0.5$ nm. Figure \ref{fig:ReducedKappa} reveals a strong
modification of the effective screening length (by factors of 2 -
10) provided $\rho_\mathrm{d}\gtrsim 1$ M and $T^*\lesssim 0.1$
(or $\epsilon\lesssim 10$). The key question is therefore if such
conditions are experimentally attainable; the answer will be
provided in the next section.

\section{Physically relevant regime}

From (\ref{fml:ReducedKappa2}) and figure \ref{fig:ReducedKappa}
we conclude that the effective screening length is significantly
larger than the Debye length, if the density of dipoles is of the
order of 1 M. We do not expect that these densities can be reached
in polar solvents (water), where $T^*$ is high, and hence $K$ is
high, such that free ions dominate according to
(\ref{fml:RhoRatio}) and figure \ref{fig:Kconstant4}. The
equilibrium constant $K$ can be very low in low-polar solvents
(oil), such that there are many more Bjerrum-pairs than free ions.
However, free ions also dissolve worse in oil than in water, and
the question is whether or not enough ionic strength is possible
to produce the Bjerrum pairs at all. In order to get an estimation
of the parameter regime where inflation of the screening length
could take place, we calculate the minimal free ion density
$\rho_\mathrm{s}^\mathrm{min}\sigma^3=\eta_\mathrm{s}^\mathrm{min}$
at which $\kappa/\bar{\kappa}>1$ to a significant degree. From
(\ref{fml:ReactionConstantBjerrum}) it can be found that $\alpha
\kappa^2 \gtrsim 1$ implies
\begin{equation}\label{fml:KappaEstimate1}
\eta_\mathrm{s}^\mathrm{min}=\sqrt{\frac{3K T^*}{4\pi}},
\end{equation}
where the dimensionless $\eta_\mathrm{s},T^*$, and $K$ were
defined in the previous section. This condition provides a lower
bound for the ion concentrations in oil where a significant effect
from pairs can be expected. If we consider the oil to be in
contact with a water reservoir, and assume the ions to partition
between the two phases due to a difference in Born self-energy
\cite{Born,Zwanikken3,Levin,Bier}, we can estimate the maximum
ionic strength $\rho_\mathrm{s}^\mathrm{max}$ in the oil. Given
the maximum ionic strength in water, being about 10 M, we find
$\rho_\mathrm{s}^\mathrm{max}= 10\cdot
\exp\bigg(\frac{\lambda_\mathrm{B}^\mathrm{water}}{\sigma}-\frac1{T^*}\bigg)$.
A significant increase of $\bar{\kappa}^{-1}$ due to Bjerrum pairs
thus requires the existence of a regime where
$\rho_\mathrm{s}^\mathrm{min}<\rho_\mathrm{s}<\rho_\mathrm{s}^\mathrm{max}$.
From a numerical analysis, however, we find that
$\rho_\mathrm{s}^\mathrm{max}<\rho_\mathrm{s}^\mathrm{min}$ for
all $T^*$ using reasonable values for $\sigma$. In other words,
the required high density of dipoles cannot be reached according
to the equilibrium constant $K$ defined in (\ref{fml:EqConst}).

In order to make some further progress, we treat solvation effects
in a slightly less naive fashion. So far we considered the ions to
have an effective radius $a_\pm$ of a few \AA, connected to a Born
self energy of several tens of $k_\mathrm{B}T$ in oil and less
than 1 $k_\mathrm{B}T$ in water. The bare radius of $\sim1$ \AA$\
$for small ions such as Na$^+$ or Cl$^-$ would overestimate the
self energies and underestimate the solvation of the ions in low
polar solvents as found for example in experiments
\cite{Leunissen}. The actual effective radius is thus larger due
to hydration shells of water molecules that form a cage around the
ion \cite{Israelachvili}. By assuming now that two ions can
approach each other up to the {\sl bare} diameter $\xi \sigma$ of
the ions, where $0<\xi<1$, in stead of $\sigma$, the effective
diameter of the ions, the equilibrium constant $K$ is found to be
much lower. This is visualized by the breaking and forming of the
structure of water molecules around the ions. Within this
speculative picture the equilibrium constant of
(\ref{fml:EqConst}) is redefined by
\begin{equation}
K_\xi^{-1} = 4\pi \Big(\frac{1}{\xi T^*}\Big{)}^3 \int
\limits_{\xi T^*}^1 \ud l\ l^2\ \exp\bigg(\frac1{l}\bigg),
\end{equation}
where an explicit energetic and entropic cost of restructuring the
layer of surrounding water molecules is ignored. A small $\xi$ can
lead to much higher densities of dipoles, by orders of magnitude
compared to $\xi = 1$. Intuitively one could expect higher order
clusters to form at such densities. We expect however that higher
order clusters are increasingly unfavourable. The energy gain by
electrostatic arguments has to compensate for both the loss of
entropy, and the energy needed to restructure the surrounding
water molecules \cite{Fennell}.

\begin{figure}[!ht] \centering
\caption{\small The equilibrium constant $K_\xi$ related to the
dimensionless temperature $T^*=\sigma/\lambda_B$ (a), and the
densities at which $\alpha \kappa^2=1$ (b), for several values of
$\xi$. The right vertical axis denotes the molar density for the
choice $\sigma =0.5$ nm.} \label{fig:Kconstant4extra3}
\subfloat[The equilibrium constant $K_\xi$ related to the
dimensionless temperature $T^*=\sigma/\lambda_B$ for several
values of $\xi$. When one focusses on the left vertical axis, the
lines also mark the points where the density of paired ions equals
the density of free cations/anions for given concentration of free
ions (see right vertical axis), and given an ionic diameter of
$\sigma=0.5$ nm. For decreasing $\xi$ the parameter regime where
the dipoles dominate increases.]{\includegraphics[width =
7cm]{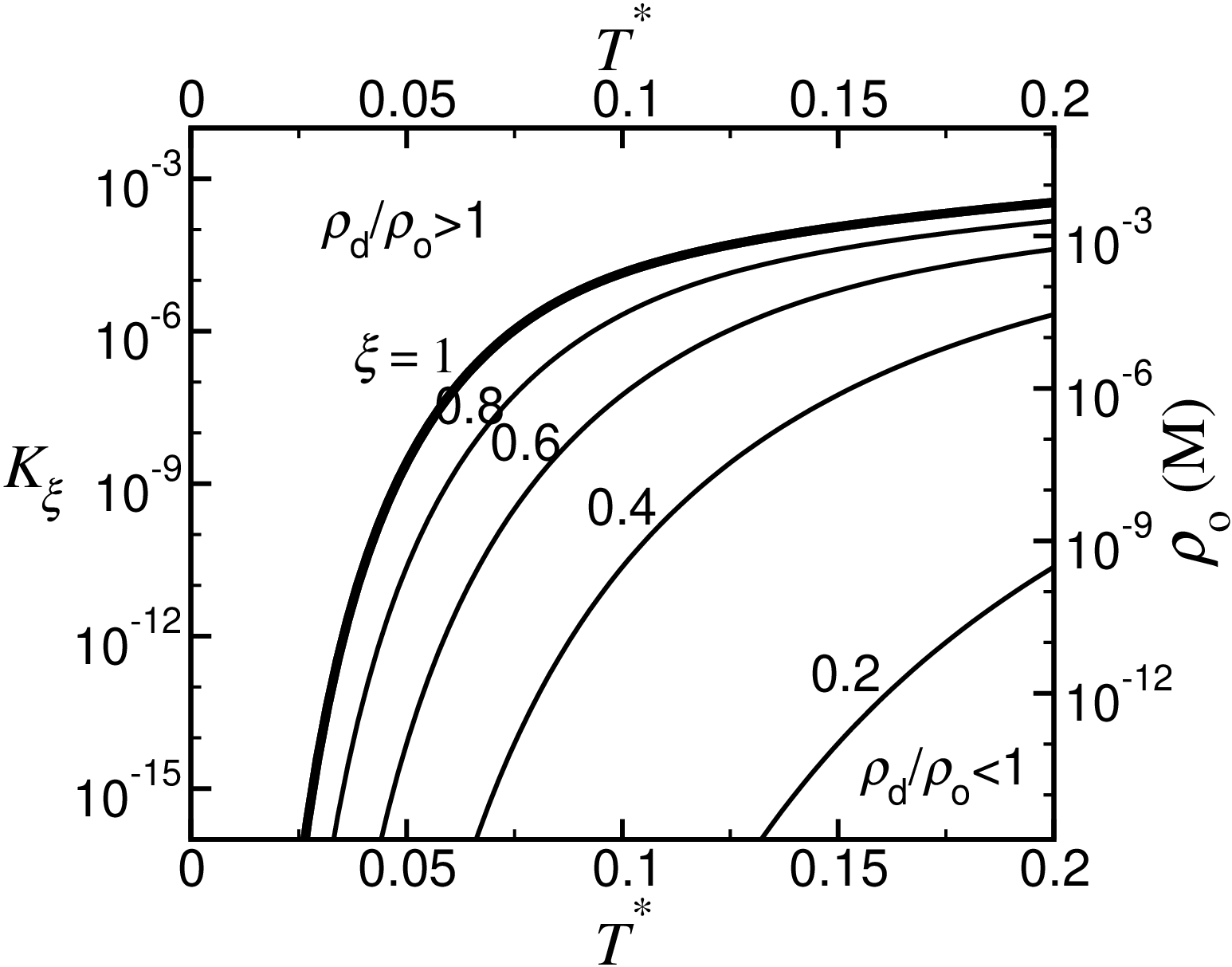}}\hspace{4mm} \subfloat[The ion densities at
which $\alpha \kappa^2=1$, i.e. where the effective screening
length $\bar{\kappa}^{-2} = 2 \kappa^{-1}$. For small $\xi$, i.e.
a small bare radius compared to the effective radius, the
screening length is increased by the presence of the dipoles,
already at low salt concentrations. The red dashed line denotes
the ion density at which the salt concentration in water is 10 M,
i.e. higher salt concentrations are not physical in
oil.]{\includegraphics[width = 7cm]{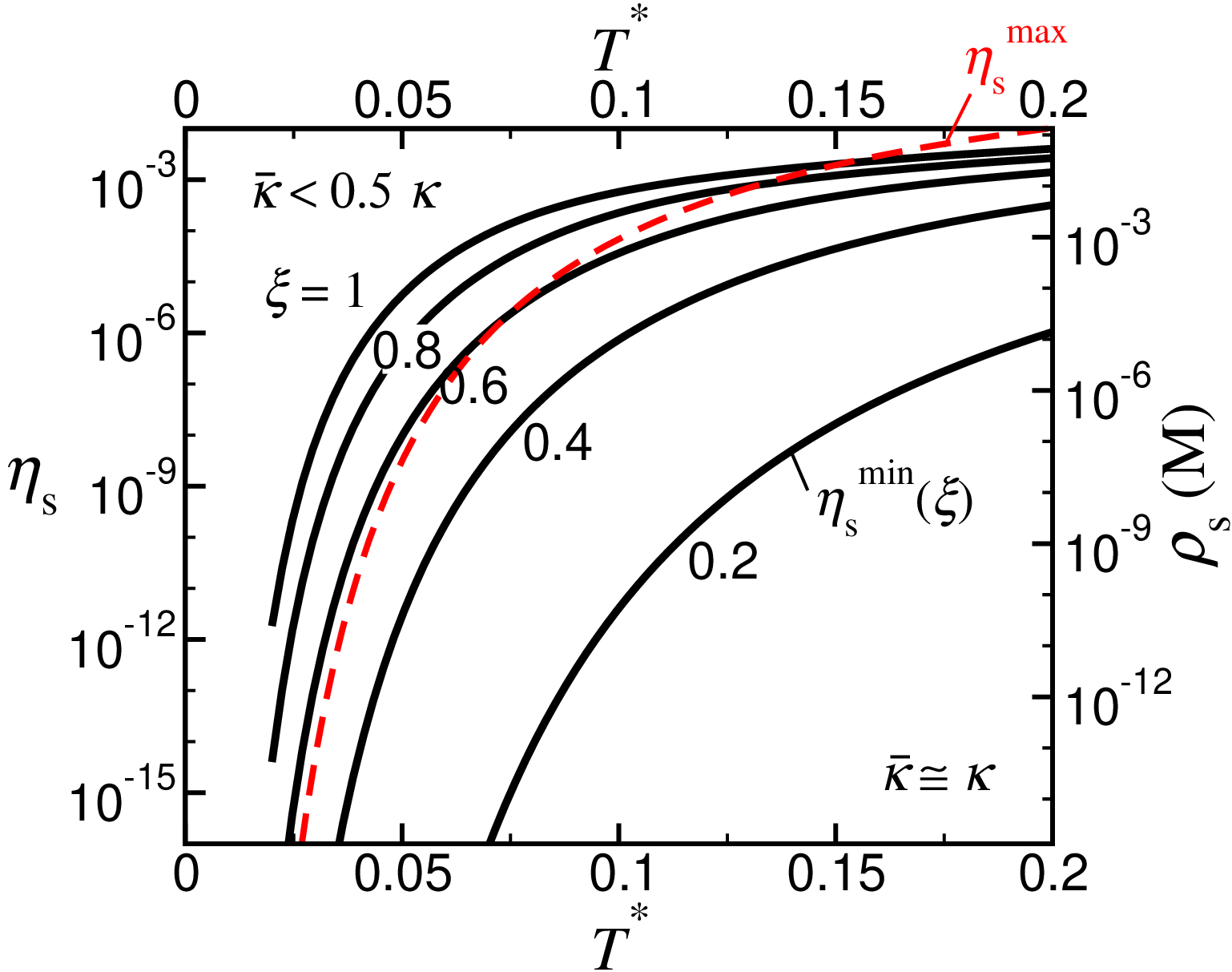}}
\end{figure}

In figure \ref{fig:Kconstant4extra3}a we plot $K$ as a function of
$T^*$ for several $\xi$; for $\xi=1$ the curve is identical to the
curve of figure \ref{fig:Kconstant4}. As in figure
\ref{fig:Kconstant4}, also the present curves separate the
high-$T^*$ regime dominated by free ions from a low-$T^*$ regime
dominated by dipoles. Reducing the contact distance from $\sigma$
to $\xi\sigma$ is immediately seen to reduce the free ion regime:
dipoles form already at higher $T^*$. The right vertical axis
denotes, again in analogy with figure \ref{fig:Kconstant4}, the
free ion concentration at the crossover
$\rho_\mathrm{s}=\rho_\mathrm{d}$ (using $\sigma=0.5$ nm to
convert to molar concentrations). Figure
\ref{fig:Kconstant4extra3}a thus reveals a lowering of the free
ion concentration at which dipole formation sets in by orders of
magnitude when $\xi$ is reduced from 1 to 0.2. Figure
\ref{fig:Kconstant4extra3}b shows the (dimensionless and molar)
maximum and minimum free ion concentrations
$\rho_\mathrm{s}^\mathrm{max}$ and
$\rho_\mathrm{s}^\mathrm{min}(\xi)$, respectively, as a function
of $T^*$ and several $\xi$. Consistent with our earlier
observations we see that $\rho_\mathrm{s}^\mathrm{min} >
\rho_\mathrm{s}^\mathrm{max}$ for $\xi\gtrsim 0.6$ in the
$T^*$-regime of interest. Interestingly, for $\xi\lesssim0.6$ a
physically attainable regime of
$\rho_\mathrm{s}^\mathrm{min}<\rho_\mathrm{s}<\rho_\mathrm{s}^\mathrm{max}$
opens up in which significant dipolar effects are to be expected
to increase the effective screening length beyond the bare one.
For the choice of $\sigma=0.5$ nm used in figure
\ref{fig:Kconstant4extra3}b, this implies that Bjerrum pairs play
an important role provided the bare ion diameter
$\xi\sigma\lesssim0.3$ nm. This seems physically reasonable.

\section{Possible observations}

We will now consider the system of \cite{Leunissen}, consisting of
micrometer-sized, strongly hydrophobic PMMA particles, dispersed
in an oily mixture of cyclohexylbromide and cis-decalin in contact
with water, containing monovalent ions. A densely packed monolayer
of colloidal particles was observed at the oil-water interface,
and a dilute bulk crystal separated by a large colloid-free zone
of $\sim100$ $\mu$m between the bulk crystal and the interface.
The system was theoretically described in \cite{Zwanikken} with a
model that will be extend here by the introduction Bjerrum pairs.
We consider strongly hydrophobic colloidal particles in oil near a
planar oil-water interface, in the presence of monovalent ions. We
focus on the distribution of particles in the direction
perpendicular to the interface. By employing the framework of
density functional theory we write the grand-potential as a
functional $\Omega[\rho,\rho_+,\rho_-,\rho_\mathrm{d}]$ of the
variational density profiles of the colloidal particles
$\rho(\mathbf{r})$, the cations $\rho_+(\mathbf{r})$, the anions
$\rho_-(\mathbf{r})$, and the dipoles
$\rho_\mathrm{d}(\mathbf{r},\mathbf{s})$, with $\mathbf{s}$ the
vector of the dipole orientation. It is identically presented in
\cite{Zwanikken} except for the dipole contributions (with a
subscript d), and given by
\begin{eqnarray}\label{fml:Omega2}
\Omega&=&\sum_{\alpha=\pm}\int\!\! \ud \mathbf{r}\,
\rho_{\alpha}(\mathbf{r})
\Big(k_BT(\ln \rho_{\alpha}(\mathbf{r})\Lambda^3 -1)+V_{\alpha}(\mathbf{r})\Big)\nonumber\\
&+& \int \ud \mathbf{r}\rho(\mathbf{r})\Big(k_BT
(\ln\frac{\eta(\mathbf{r})}{\eta_0}-1) +
V(\mathbf{r})\Big)\nonumber\\&+& \int \ud \mathbf{r} \ud \mathbf{s} \rho_\mathrm{d}(\mathbf{r},\mathbf{s})\Big(k_BT (\ln \rho_\mathrm{d}(\mathbf{r},\mathbf{s})\Lambda^3-1) + V_+(\mathbf{r})+V_-(\mathbf{r})+ \Delta G(\mathbf{r})\Big)\nonumber\\
&-& \sum_{\alpha=\pm} \mu_\alpha \int\!\! \ud \mathbf{r}\ \Big(\rho_\alpha(\mathbf{r}) + \int\!\! \ud \mathbf{s} \rho_\mathrm{d}(\mathbf{r},\mathbf{s})\Big) \nonumber\\
&+&k_BT\int \ud \mathbf{r}
\Big(\rho(\mathbf{r})\Psi(\bar{\eta}(\mathbf{r}))+\frac{1}{2}Q(\mathbf{r})\phi(\mathbf{r})\Big),
\end{eqnarray}
where $\eta(\mathbf{r})=4\pi a^3\rho(\mathbf{r})/3$ is the
colloidal packing fraction, and where the first and second line
are the ideal-gas grand-potential functionals of the ions and the
colloidal particles in their external fields, respectively, the
third line is the ideal gas free energy of the dipoles and the
binding free energy, the fourth line a grand canonical
contribution (for fixed chemical potentials), and the last line
describes the hard-core and Coulomb interactions,
\cite{Zwanikken}. The chemical potentials of the colloidal
particles is represented in terms of a reference colloid packing
fraction $\eta_0$ to be discussed below. The total local charge
number density
$Q(\mathbf{r})=Z\rho(\mathbf{r})+\rho_+(\mathbf{r})-\rho_-(\mathbf{r})+\int
\ud \mathbf{s} [\rho_d(\mathbf{r} +
\mathbf{s}\frac{\bar{\sigma}}2,\mathbf{s}) -
\rho_\mathrm{d}(\mathbf{r} -
\mathbf{s}\frac{\bar{\sigma}}2,\mathbf{s})]$, with $\mathbf{s}$
the unit vector denoting the direction of the dipole, and
$\bar{\sigma}$ the mean distance between the centers of the ions
of a pair (previously found to be $\bar{\sigma}\simeq\sigma$ in
low dielectric media, see (\ref{fml:MeanSigma}) and figure
\ref{fig:rMEAN}). For small $\bar{\sigma}$ the last term in the
expression for $Q(\mathbf{r})$ reduces to
\begin{equation}\label{fml:dipolecharge}
\int \ud \mathbf{s}\ \bar{\sigma} \nabla
\rho_\mathrm{d}(\mathbf{r},\mathbf{s})\cdot \mathbf{s}.
\end{equation}
We obtain the equilibrium distribution of colloidal particles by
minimization of the functional (\ref{fml:Omega2}) with respect to
$\eta(\mathbf{r})$, which reduces to $\eta(z)$ due to the symmetry
of the system. Minimizations with respect to the densities
$\rho_\pm(z)$ yield the Boltzmann distributions for the free ions,
also given in \cite{Zwanikken} (only we use the slightly different
notation $\rho(\infty) \equiv \rho_\mathrm{s}$ here) and dipole
density $\rho_\mathrm{d}(\mathbf{r},\mathbf{s})$
\begin{equation}\label{fml:Boltzmann2}
\rho_\mathrm{d}(z)= \rho_\mathrm{d} \frac{\sinh(\bar{\sigma})
|\phi'(z)|}{\bar{\sigma} |\phi'(z)|},
\end{equation}
where the expression is integrated over $\mathbf{s}$, and where we
used the relations between $K$, $\rho_\mathrm{d}$ and
$\rho_\mathrm{s}$ given in earlier sections. Combining these with
the Poisson equation yields a Poisson-Boltzmann equation
(\ref{fml:PBAbrashkin}) from which we can find the electrostatic
potential $\phi(z)$, for the boundary conditions
\begin{eqnarray}\label{fml:PB5}
\lim_{z\uparrow0} \epsilon_w \phi'(z) &=& \lim_{z\downarrow0}
\epsilon_o \phi'(z) \hspace{3mm};\hspace{3mm}
\lim_{z\rightarrow\pm\infty} \phi'(z)=0,\nonumber
\end{eqnarray}
where $\epsilon_\mathrm{w}$ is the relative permittivity of water,
and $\epsilon_\mathrm{o}$ that of oil. The five equations of the
five unknowns (the particle densities and the electrostatic
potential) are solved numerically.

\begin{figure}[!ht]
\centering
\includegraphics[width = 7.4cm]{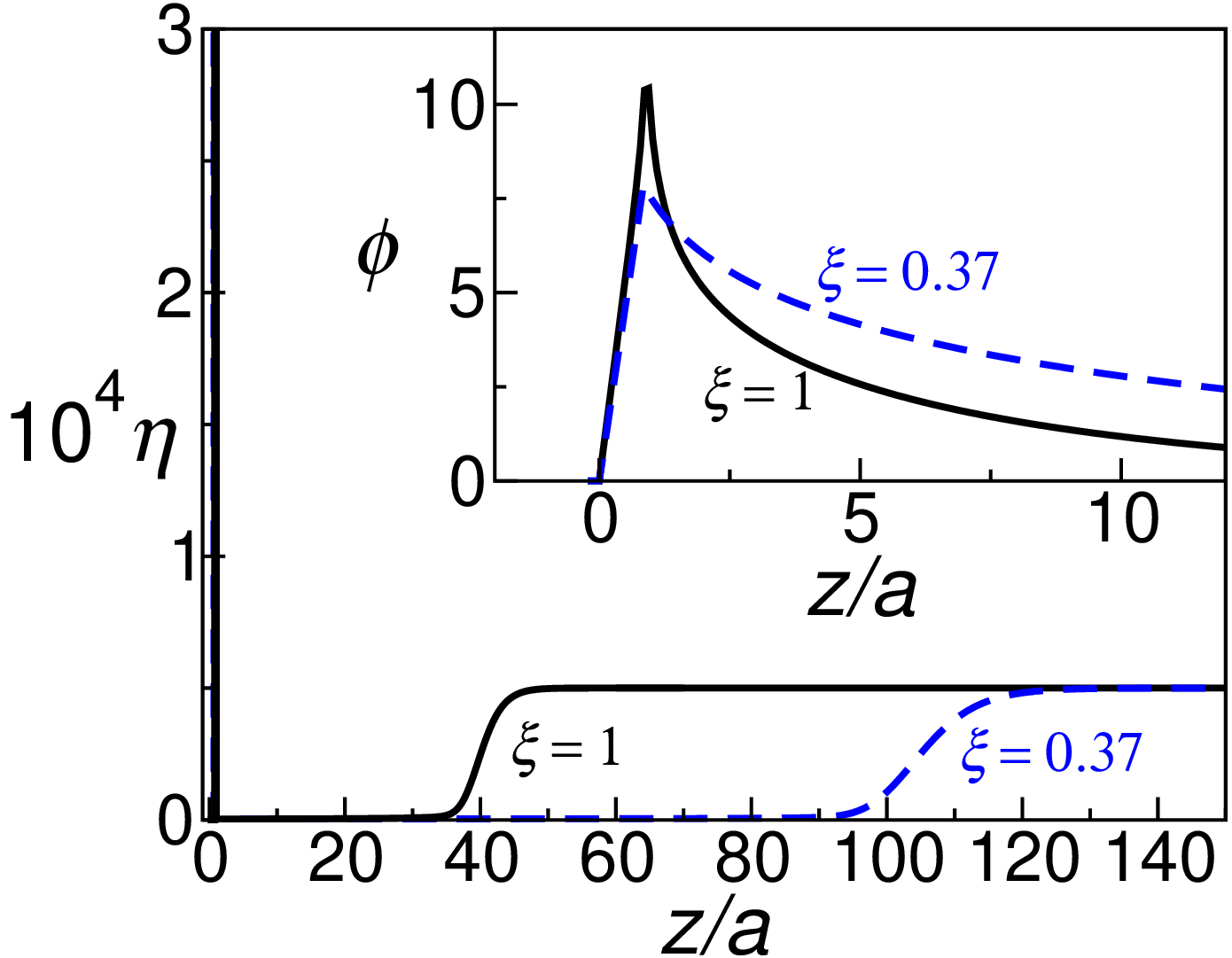}
\caption{\small The packing fraction profile $\eta(z)$ of strongly
hydrophobic, oil-dispersed colloidal spheres (radius $a=1\mu$m,
charge $Z=450$) in the vicinity of a planar interface at $z=0$
between water ($z<0$, dielectric constant $\epsilon_w=80$) and oil
$(z>0, \epsilon_o=5.2)$, for a colloidal bulk packing fraction
$\eta_b=\eta(\infty)=5\times 10^{-5}$ of weakly wetting colloidal
particles ($\cos\theta=0.987$) and screening length in oil
$\kappa^{-1}/a=8$. The curves show the influence of Bjerrum pairs
on the depletion zone. For $\xi=1$ the effect of Bjerrum pairs can
be ignored ($\bar{\kappa}^{-1}/a=\kappa^{-1}/a=8$). For $\xi=0.37$
Bjerrum pairs cannot be ignored and the screening length
effectively increases up to $\bar{\kappa}^{-1}/a=20$, due to an
effective increase of the permittivity, resulting in a long range
repulsion between the monolayer and the bulk crystal over 100
$\mu$m.} \label{fig:figure1a}
\end{figure}

Figure \ref{fig:figure1a} shows the resulting packing fraction
profile $\eta(z)$, as well as the electrostatic potential
$\phi(z)$ in the inset, for $\xi = 1$ and $\xi = 0.37$. The
parameters are the colloidal radius $a=1$ $\mu$m, the colloidal
charge $Z=450$, the relative permittivity of water
$\epsilon_\mathrm{w}=80$, and that of oil
$\epsilon_\mathrm{o}=5.2$, and also the external potentials $V(z)$
and $V\pm(z)$ are identical to those in \cite{Zwanikken}, i.e.
$V(z)$ is a Pieranski potential \cite{Pieranski} with contact
angle $\cos \theta=0.987$ and an oil water interfacial tension of
$\gamma_\mathrm{ow}=9$ mN/m, and $V_\pm(z)$ is based on Born
self-energy differences in oil and water with ionic radii $a_\pm =
0.3 $ nm (i.e. $\sigma=0.6$ nm).

The curve $\eta(z)$ for $\xi=1$ is virtually identical to the one
published in \cite{Zwanikken} (i.e. the effect of the Bjerrum
pairs can be ignored completely) and reveals a strong monolayer
adsorption at $z\simeq a$, a colloid-free zone for $1\lesssim
z/a\lesssim 30$, and a colloidal crystal \cite{shklovskii} with a
packing fraction $\eta(\infty) = \eta_\mathrm{b} = 5\cdot10^{-5}$
at $z\gtrsim 30$. Experimentally, however, a colloid-free zone
that extends to $z\simeq 100$ $\mu$m was observed \cite{Leunissen}
for these parameters. Comparing this with the curve for $\xi =
0.37$ in figure \ref{fig:figure1a} yields a much better agreement
with the experimental observation. We speculate, therefore, that
Bjerrum pairing is an interesting feature for further study in
these oily solvents.

\section{Conclusion}

In this paper we considered the effect of Bjerrum pairs on the
screening length, and concluded that it can be significantly
larger in low-polar media than the Debye length that is calculated
from the free ion concentration (for example obtained by
conductivity measurements) and the bare solvent dielectric
constant. Due to the coupling of free ions and dipoles through an
association-dissociation equilibrium, we predict the effective
screening length to scale as $\bar{\kappa}^{-1}\propto
\sqrt{\rho_\mathrm{s}}$ at relatively high salt concentrations, in
contrast to the scaling $\kappa^{-1} \propto
1/\sqrt{\rho_\mathrm{s}}$ for the Debye length, where
$\rho_\mathrm{s}$ is the free ion concentration. A large
concentration of Bjerrum pairs was found to change the dielectric
constant of the medium effectively. By a naive treatment of
solvation effects of the ions, the required dipole concentrations
seem to be unattainable for physical parameters, such that the
effect of Bjerrum pairs could be neglected completely. After
making the distinction between an {\sl effective} ionic diameter,
due to hydration shells that lower the self-energy, and a {\sl
bare} ionic diameter, determining the closest distance between two
ions, a regime of physical parameters was found where inflation of
the screening length could be expected. Our results provide a
possible explanation for the extremely large colloid-free zone
that was observed in recent experiments \cite{Leunissen}. Clearly,
however, more research is needed to investigate this effect.

\paragraph{Acknowledgements} We would like to thank Chantal
Valeriani, Philip Camp, and Marjolein Dijkstra for stimulating
discussions.

\end{document}